\title{Extension of the HF program to partially filled f-subshells}
\author{ Gediminas Gaigalas\\
        {\em Institute of Theoretical Physics and Astronomy,}\\
        {\em A. Go\v{s}tauto 12, Vilnius 2600, LITHUANIA}\\ \
        \\
Charlotte Froese Fischer \\
        {\em Department of Computer Science, Box 6035 B,}\\
        {\em Vanderbilt University, Nashville, TN 37235, USA}}
\documentstyle[11pt]{article}
\begin{document}
\maketitle

A new version of a Hartree-Fock program is presented that
includes extensions for partially filled $f$-subshells. The program
allows the calculation of term dependent
Hartree-Fock orbitals and energies in $LS$ coupling
for configurations with no more than two open subshells,
including $f$-subshells.

\twocolumn

\centerline{\large \bf PROGRAM SUMMARY}
\medskip

\begin{flushleft}
{\em Title of program:} HF96
\vskip 0.15in

{\em Catalogue number:} ADDZ
\vskip 0.15in

{\em Program obtainable from:} CPC Program Library, Queen's
University of Belfast, N. Ireland (see application form in this issue)
\vskip 0.15in

{\em Licensing provisions:} none
\vskip 0.15in

{\em Computer:} SUN Sparcstation 1
\vskip 0.15in

{\em Installation:} Vanderbilt University, Nashville, TN 37235, USA
\vskip 0.15in

{\em Operating System under which the program is executed:} Sun UNIX
\vskip 0.15in

{\em Programming language used:} FORTRAN 77
\vskip 0.15in

{\em Memory required to execute with typical data:} 449K Bytes
\vskip 0.15in

{\em No. of bits in a word:}  32
\vskip 0.15in

{\em Peripherals used:} terminal, disk
\vskip 0.15in

{\em No. of lines in combined program and test deck:} 4703
\vskip 0.15in

{\em CPC program Library subprograms used:} none
\vskip 0.15in

\vskip 0.15in

{\em Nature of physical problem}\\
This program determines non-relativistic, numerical radial functions and
computes energy levels in the Hartree-Fock approximation.
\vskip 0.15in

{\em Method of solution}\\
The new version of the program retains the design and structure
of the previous one~[1]. The same self-consistent field method is used
to solve the differential equation~[2]. Tables from Ref.[3] that
define deviations of term energies from the average energy of
the configuration $E^{av}$
have been extended to include the partially filled $f$-subshells.
\vskip 0.15in

{\em Restriction on the complexity of the program}\\
Only one or two open subshells are allowed. Any number of $s$, $p$, $d$ and $f$
electrons are allowed in the case of  one open subshell.
In the two open subshell case, the program can calculate the term
dependent energy expression only for the following
configurations:
\begin{enumerate}
\item $(ns)(n'l)^{N}$ ,where $l=0,1,2, 3$,

\item $(np)^{N}(n'l)$ ,where $l=0,1,2,3,...$~~,

\item $(nf)(n'd)$.
\end{enumerate}

\vskip 0.15in

{\em Typical running time}\\
The time required to execute the test case on a SUN Sparcstation~1
was 64.5 seconds.
\vskip 0.15in

{\em References}
\begin{enumerate}
\item C. Froese Fischer, Comput. Phys. Commun. 43 (1987) 355.

\item C.Froese Fischer, The Hartree-Fock Method for Atoms:A Numerical
Approach (Wiley, New York, 1977).

\item John C. Slater, Quantum Theory of Atomic Structure, Vol II.
(McGraw-Hill Book Company, New York, 1960).
\end{enumerate}
\end{flushleft}

\onecolumn

\section*{LONG WRITE-UP}

\section{Introduction}

Recently there has been a tremendous growth of interest in lanthanide
and actinide chemistry and physics, yet relatively few
calculations have been performed for these sytems.  Though accurate
results will require a fully relativistic treatment in the
$jj$-coupling scheme, that also accounts for the
nuclear effects on the energy level structure as is done in
GRASP92~\cite{GRASP}, for example, much can be learned about
quantitative behaviour of atomic properties from simpler non-relativistic
calculations. For example, Tatewaki {\sl et al}~\cite{Tatewaki}
recently found
that the non-relativistic Hartree-Fock ionization potentials of singly
ionized lanthanides follow the observed trends,

The numerical Hartree-Fock program, HF86, published
earlier~\cite{FF} could only treat configurations with open
$f$-subshells containing more than one electron, in the {\sc
average-energy-of-configuration} approximation~\cite{S}.  In the present
version, HF96, this restriction is removed, though like the
earlier version, it is only able to compute the term-dependent
orbitals when an expression for the energy can be determined from
relatively simple formulas.  More complex configuration states
will still need the full power of the angular momentum programs
part of the MCHF Atomic Structure Package~\cite{atsp}, extended also for
open $f$-subshells.

\section{Program organization}

The new version of the HF program has the same design as the previous
one~\cite{FF}.
It gets an expression for the average energy from the occupation
numbers for the subshells and then,
corrects the average energy expression with a
deviation so that
$$E(LS)=E^{av}+\Delta E(LS).$$
The deviation, $\Delta E(LS)$, has the property that the expression is the same for particles
and holes. Thus one needs to be concerned only with up to half-filled
subshells and their deviations.
There are
simple rules for other cases (for details see~\cite{S}).

The HF program~\cite{FF} is extended to include the cases:
\begin{enumerate}
\item  $(nf)^{N}$,
\item  $(nf)^{N}(n's)$,
\item  $(nf)(n'd)$,
\end{enumerate}
Input data for the new version of HF is the same as before.
One needs only to take into account that the  classification of
terms of the $f$ subshell
is more complicated than for $s$, $p$, $d$ subshells.
It is related to the fact that for the classification of $f$-subshell states
the characteristics $S'$ (multiplicity), $L$ (total orbital
momentum), and $\nu$ (seniority) are not enough.
We use a classification
of terms similar to that used by Nielson and Koster~\cite{NK}.
Namely for a particular subshell instead of $\nu$, when the term
is not unique, a single character ``number'', $Nr$, is used. Its value is found in
Table~1 where all terms for $f$-subshells are presented.  In most
cases, $Nr$ appears to be a digit, but since it is a single
character, the single letter $A$ is used instead of the number
10.

In many instances, the term can be uniquely determined and
neither seniority $\nu$ nor $Nr$ are need.  However, when more
than one coupling is possible, additional information is
requested.
Consider a calculation for the $^4F$ term of the
configuration {\tt 4f(4)6s(1)} for which the coupling of 4f(4) could
be either $^3F$ or $^5F$.  Table~1 shows there is one $^5F$ term
and four $^3F$ terms.  In cases of such ambiguity, the user will
need to provide additional information for the term and $Nr$ of
the ``parent'', or more precisely the {\tt l**n} subshell.  This
is illustrated in the {\bf Test Run Output}.

\section{Example}

Our example shows the input to the interactive Hartree Fock
program and some of the intermediate output.  The calculation is
for the total energy of the level
$1s^{2}2s^{2}2p^{6}3s^{2}3p^{6}4s^{2}4p^{6}4d^{10}5s^{2}5p^{6}4f^{4}
~^{3}F^{4}_{W=(211)U=(21)} 6s^{1}~^{4}F$
of $Nd^{+}$. Initially, calculations were requested for $^4F$. Then
the program determined an {\tt Ambiguous l**n parent term} and
requested information about the term and $Nr$ for 4f(4). A
calculation was desired for the $^3F$ parent with $Nr$ =3.  This
information was entered as a three-character string, {\tt 3F3}.
With this information, the energy expression could be
determined.

\begin{table}
\caption{Classification of  $f$- subshell states.}
% Lentele 1.1
\begin{center}
\begin{tabular}{ l c c c l| r c c c l| r c c c l }
\hline \hline
$LS$  & $\nu$ & $W$ & $U$ & $Nr$ &
$LS$  & $\nu$ & $W$ & $U$ & $Nr$ &
$LS$  & $\nu$ & $W$ & $U$ & $Nr$\\ \hline
\multicolumn{4}{l}{\sl subshell $f^{1}$} & &
\multicolumn{4}{l}{\sl subshell $f^{4}$} & &
$^{1}S$  &  $0$ & $(000)$ & $(00)$ & $1$  \\
$^{2}F$  &  $1$ & $(100)$ & $(10)$ & $1$ &
$^{5}S$  &  $4$ & $(111)$ & $(00)$ & $0$ &
$^{1}S$  &  $4$ & $(220)$ & $(22)$ & $2$  \\
\multicolumn{4}{l}{\sl subshell  $f^{2}$} & &
$^{5}D$  &  $4$ & $(111)$ & $(20)$ & $1$ &
$^{1}D$  &  $2$ & $(200)$ & $(20)$ & $1$  \\
$^{3}P$  &  $2$ & $(110)$ & $(11)$ & $1$ &
$^{5}F$  &  $4$ & $(111)$ & $(10)$ & $1$ &
$^{1}D$  &  $4$ & $(220)$ & $(20)$ & $2$  \\
$^{3}F$  &  $2$ & $(110)$ & $(10)$ & $1$ &
$^{5}G$  &  $4$ & $(111)$ & $(20)$ & $1$ &
$^{1}D$  &  $4$ & $(220)$ & $(21)$ & $3$  \\
$^{3}H$  &  $2$ & $(110)$ & $(11)$ & $1$ &
$^{5}I$  &  $4$ & $(111)$ & $(20)$ & $1$ &
$^{1}D$  &  $4$ & $(220)$ & $(22)$ & $4$  \\
$^{1}S$  &  $0$ & $(000)$ & $(00)$ & $1$ &
$^{3}P$  &  $2$ & $(110)$ & $(11)$ & $1$ &
$^{1}F$  &  $4$ & $(220)$ & $(21)$ & $1$  \\
$^{1}D$  &  $2$ & $(200)$ & $(20)$ & $1$ &
$^{3}P$  &  $4$ & $(211)$ & $(11)$ & $2$ &
$^{1}G$  &  $2$ & $(200)$ & $(20)$ & $1$  \\
$^{1}G$  &  $2$ & $(200)$ & $(20)$ & $1$ &
$^{3}P$  &  $4$ & $(211)$ & $(30)$ & $3$ &
$^{1}G$  &  $4$ & $(220)$ & $(20)$ & $2$  \\
$^{1}I$  &  $2$ & $(200)$ & $(20)$ & $1$ &
$^{3}D$  &  $4$ & $(211)$ & $(20)$ & $1$ &
$^{1}G$  &  $4$ & $(220)$ & $(21)$ & $3$  \\
\multicolumn{4}{l}{\sl subshell $f^{3}$}& &
$^{3}D$  &  $4$ & $(211)$ & $(21)$ & $2$ &
$^{1}G$  &  $4$ & $(220)$ & $(22)$ & $4$  \\
$^{4}S$  &  $3$ & $(111)$ & $(00)$ & $1$ &
$^{3}F$  &  $2$ & $(110)$ & $(10)$ & $1$ &
$^{1}H$  &  $4$ & $(220)$ & $(21)$ & $1$  \\
$^{4}D$  &  $3$ & $(111)$ & $(20)$ & $1$ &
$^{3}F$  &  $4$ & $(211)$ & $(10)$ & $2$ &
$^{1}H$  &  $4$ & $(220)$ & $(22)$ & $2$  \\
$^{4}F$  &  $3$ & $(111)$ & $(10)$ & $1$ &
$^{3}F$  &  $4$ & $(211)$ & $(21)$ & $3$ &
$^{1}I$  &  $2$ & $(200)$ & $(20)$ & $1$  \\
$^{4}G$  &  $3$ & $(111)$ & $(20)$ & $1$ &
$^{3}F$  &  $4$ & $(211)$ & $(30)$ & $4$ &
$^{1}I$  &  $4$ & $(220)$ & $(20)$ & $2$  \\
$^{4}I$  &  $3$ & $(111)$ & $(20)$ & $1$ &
$^{3}G$  &  $4$ & $(211)$ & $(20)$ & $1$ &
$^{1}I$  &  $4$ & $(220)$ & $(22)$ & $3$  \\
$^{2}P$  &  $3$ & $(210)$ & $(11)$ & $1$ &
$^{3}G$  &  $4$ & $(211)$ & $(21)$ & $2$ &
$^{1}K$  &  $4$ & $(220)$ & $(21)$ & $1$  \\
$^{2}D$  &  $3$ & $(210)$ & $(20)$ & $1$ &
$^{3}G$  &  $4$ & $(211)$ & $(30)$ & $3$ &
$^{1}L$  &  $4$ & $(220)$ & $(21)$ & $1$  \\
$^{2}D$  &  $3$ & $(210)$ & $(21)$ & $2$ &
$^{3}H$  &  $2$ & $(110)$ & $(11)$ & $1$ &
$^{1}L$  &  $4$ & $(220)$ & $(22)$ & $2$  \\
$^{2}F$  &  $1$ & $(100)$ & $(10)$ & $1$ &
$^{3}H$  &  $4$ & $(211)$ & $(11)$ & $2$ &
$^{1}N$  &  $4$ & $(220)$ & $(22)$ & $1$  \\
$^{2}F$  &  $3$ & $(210)$ & $(21)$ & $2$ &
$^{3}H$  &  $4$ & $(211)$ & $(21)$ & $3$ &
\multicolumn{5}{l}{\sl subshell $f^{5}$} \\
$^{2}G$  &  $3$ & $(210)$ & $(20)$ & $1$ &
$^{3}H$  &  $4$ & $(211)$ & $(30)$ & $4$ &
$^{6}P$  &  $5$ & $(110)$ & $(11)$ & $0$  \\
$^{2}G$  &  $3$ & $(210)$ & $(21)$ & $2$ &
$^{3}I$  &  $4$ & $(211)$ & $(20)$ & $1$ &
$^{6}F$  &  $5$ & $(110)$ & $(10)$ & $0$  \\
$^{2}H$  &  $3$ & $(210)$ & $(11)$ & $1$ &
$^{3}I$  &  $4$ & $(211)$ & $(30)$ & $2$ &
$^{6}H$  &  $5$ & $(110)$ & $(11)$ & $0$  \\
$^{2}H$  &  $3$ & $(210)$ & $(21)$ & $2$ &
$^{3}K$  &  $4$ & $(211)$ & $(21)$ & $1$ &
$^{4}S$  &  $3$ & $(111)$ & $(00)$ & $1$  \\
$^{2}I$  &  $3$ & $(210)$ & $(20)$ & $1$ &
$^{3}K$  &  $4$ & $(211)$ & $(30)$ & $2$ &
$^{4}P$  &  $5$ & $(211)$ & $(11)$ & $1$  \\
$^{2}K$  &  $3$ & $(210)$ & $(21)$ & $1$ &
$^{3}L$  &  $4$ & $(211)$ & $(21)$ & $1$ &
$^{4}P$  &  $5$ & $(211)$ & $(30)$ & $2$  \\
$^{2}L$  &  $3$ & $(210)$ & $(21)$ & $1$ &
$^{3}M$  &  $4$ & $(211)$ & $(30)$ & $1$ &
$^{4}D$  &  $3$ & $(111)$ & $(20)$ & $1$  \\
\hline

\end{tabular}
\end{center}
\end{table}

\clearpage

% Lentele 1.2

\begin{table}
%\caption{Lenteles~1 (tesinys)}
\begin{center}
\begin{tabular}{ l c c c l| r c c c l| r c c c l }
\hline \hline
$LS$  & $\nu$ & $W$ & $U$ & $Nr$ &
$LS$  & $\nu$ & $W$ & $U$ & $Nr$ &
$LS$  & $\nu$ & $W$ & $U$ & $Nr$\\ \hline
$^{4}D$  &  $5$ & $(211)$ & $(20)$ & $2$ &
$^{2}D$  &  $5$ & $(221)$ & $(31)$ & $5$ &
$^{2}K$  &  $5$ & $(221)$ & $(30)$ & $3$  \\
$^{4}D$  &  $5$ & $(211)$ & $(21)$ & $3$ &
$^{2}F$  &  $1$ & $(100)$ & $(10)$ & $1$ &
$^{2}K$  &  $5$ & $(221)$ & $(31)$ & $4$  \\
$^{4}F$  &  $3$ & $(111)$ & $(10)$ & $1$ &
$^{2}F$  &  $3$ & $(210)$ & $(21)$ & $2$ &
$^{2}K$  &  $5$ & $(221)$ & $(31)$ & $5$  \\
$^{4}F$  &  $5$ & $(211)$ & $(10)$ & $2$ &
$^{2}F$  &  $5$ & $(221)$ & $(10)$ & $3$ &
$^{2}L$  &  $3$ & $(210)$ & $(21)$ & $3$  \\
$^{4}F$  &  $5$ & $(211)$ & $(21)$ & $3$ &
$^{2}F$  &  $5$ & $(221)$ & $(21)$ & $4$ &
$^{2}L$  &  $5$ & $(221)$ & $(21)$ & $2$  \\
$^{4}F$  &  $5$ & $(211)$ & $(30)$ & $4$ &
$^{2}F$  &  $5$ & $(221)$ & $(30)$ & $5$ &
$^{2}L$  &  $5$ & $(221)$ & $(31)$ & $5$  \\
$^{4}G$  &  $3$ & $(111)$ & $(20)$ & $1$ &
$^{2}F$  &  $5$ & $(221)$ & $(31)$ & $6$ &
$^{2}M$  &  $5$ & $(221)$ & $(30)$ & $1$  \\
$^{4}G$  &  $5$ & $(211)$ & $(20)$ & $2$ &
$^{2}F$  &  $5$ & $(221)$ & $(31)$ & $7$ &
$^{2}M$  &  $5$ & $(221)$ & $(31)$ & $2$  \\
$^{4}G$  &  $5$ & $(211)$ & $(21)$ & $3$ &
$^{2}G$  &  $3$ & $(210)$ & $(20)$ & $1$ &
$^{2}N$  &  $5$ & $(221)$ & $(31)$ & $1$  \\
$^{4}G$  &  $5$ & $(211)$ & $(30)$ & $4$ &
$^{2}G$  &  $3$ & $(210)$ & $(21)$ & $2$ &
$^{2}O$  &  $5$ & $(221)$ & $(31)$ & $0$  \\
$^{4}H$  &  $5$ & $(211)$ & $(11)$ & $1$ &
$^{2}G$  &  $5$ & $(221)$ & $(20)$ & $3$ &
\multicolumn{5}{l}{\sl subshell $f^{6}$}  \\
$^{4}H$  &  $5$ & $(211)$ & $(21)$ & $2$ &
$^{2}G$  &  $5$ & $(221)$ & $(21)$ & $4$ &
$^{7}F$  &  $6$ & $(100)$ & $(10)$ & $0$  \\
$^{4}H$  &  $5$ & $(211)$ & $(30)$ & $3$ &
$^{2}G$  &  $5$ & $(221)$ & $(30)$ & $5$ &
$^{5}S$  &  $4$ & $(111)$ & $(00)$ & $0$  \\
$^{4}I$  &  $3$ & $(111)$ & $(20)$ & $1$ &
$^{2}G$  &  $5$ & $(221)$ & $(31)$ & $6$ &
$^{5}P$  &  $6$ & $(210)$ & $(11)$ & $0$  \\
$^{4}I$  &  $5$ & $(211)$ & $(20)$ & $2$ &
$^{2}H$  &  $3$ & $(210)$ & $(11)$ & $1$ &
$^{5}D$  &  $4$ & $(111)$ & $(20)$ & $1$  \\
$^{4}I$  &  $5$ & $(211)$ & $(30)$ & $3$ &
$^{2}H$  &  $3$ & $(210)$ & $(21)$ & $2$ &
$^{5}D$  &  $6$ & $(210)$ & $(20)$ & $2$  \\
$^{4}K$  &  $5$ & $(211)$ & $(21)$ & $1$ &
$^{2}H$  &  $5$ & $(221)$ & $(11)$ & $3$ &
$^{5}D$  &  $6$ & $(210)$ & $(21)$ & $3$  \\
$^{4}K$  &  $5$ & $(211)$ & $(30)$ & $2$ &
$^{2}H$  &  $5$ & $(221)$ & $(21)$ & $4$ &
$^{5}F$  &  $4$ & $(111)$ & $(10)$ & $1$  \\
$^{4}L$  &  $5$ & $(211)$ & $(21)$ & $1$ &
$^{2}H$  &  $5$ & $(221)$ & $(30)$ & $5$ &
$^{5}F$  &  $6$ & $(210)$ & $(21)$ & $2$  \\
$^{4}M$  &  $5$ & $(211)$ & $(30)$ & $0$ &
$^{2}H$  &  $5$ & $(221)$ & $(31)$ & $6$ &
$^{5}G$  &  $4$ & $(111)$ & $(20)$ & $1$  \\
$^{2}P$  &  $3$ & $(210)$ & $(11)$ & $1$ &
$^{2}H$  &  $5$ & $(221)$ & $(31)$ & $7$ &
$^{5}G$  &  $6$ & $(210)$ & $(20)$ & $2$  \\
$^{2}P$  &  $5$ & $(221)$ & $(11)$ & $2$ &
$^{2}I$  &  $3$ & $(210)$ & $(20)$ & $1$ &
$^{5}G$  &  $6$ & $(210)$ & $(21)$ & $3$  \\
$^{2}P$  &  $5$ & $(221)$ & $(30)$ & $3$ &
$^{2}I$  &  $5$ & $(221)$ & $(20)$ & $2$ &
$^{5}H$  &  $6$ & $(210)$ & $(11)$ & $1$  \\
$^{2}P$  &  $5$ & $(221)$ & $(31)$ & $4$ &
$^{2}I$  &  $5$ & $(221)$ & $(30)$ & $3$ &
$^{5}H$  &  $6$ & $(210)$ & $(21)$ & $2$  \\
$^{2}D$  &  $3$ & $(210)$ & $(20)$ & $1$ &
$^{2}I$  &  $5$ & $(211)$ & $(31)$ & $4$ &
$^{5}I$  &  $4$ & $(111)$ & $(20)$ & $1$  \\
$^{2}D$  &  $3$ & $(210)$ & $(21)$ & $2$ &
$^{2}I$  &  $5$ & $(221)$ & $(31)$ & $5$ &
$^{5}I$  &  $6$ & $(210)$ & $(20)$ & $2$  \\
$^{2}D$  &  $5$ & $(221)$ & $(20)$ & $3$ &
$^{2}K$  &  $3$ & $(210)$ & $(21)$ & $1$ &
$^{5}K$  &  $6$ & $(210)$ & $(21)$ & $0$  \\
$^{2}D$  &  $5$ & $(221)$ & $(21)$ & $4$ &
$^{2}K$  &  $5$ & $(221)$ & $(21)$ & $2$ &
$^{5}L$  &  $6$ & $(210)$ & $(21)$ & $0$  \\
\hline

\end{tabular}
\end{center}
\end{table}

\clearpage
% Lentele 1.3

\begin{table}
%\caption{Termu charakteristikos $f$- sluoksniui.}
\begin{center}
\begin{tabular}{ l c c c l| r c c c l| r c c c l }
\hline \hline
$LS$  & $\nu$ & $W$ & $U$ & $Nr$ &
$LS$  & $\nu$ & $W$ & $U$ & $Nr$ &
$LS$  & $\nu$ & $W$ & $U$ & $Nr$\\ \hline
$^{3}P$  &  $2$ & $(110)$ & $(11)$ & $1$ &
$^{3}H$  &  $4$ & $(211)$ & $(11)$ & $2$ &
$^{1}S$  &  $0$ & $(000)$ & $(00)$ & $1$  \\
$^{3}P$  &  $4$ & $(211)$ & $(11)$ & $2$ &
$^{3}H$  &  $4$ & $(211)$ & $(21)$ & $3$ &
$^{1}S$  &  $4$ & $(220)$ & $(22)$ & $2$  \\
$^{3}P$  &  $4$ & $(211)$ & $(30)$ & $3$ &
$^{3}H$  &  $4$ & $(211)$ & $(30)$ & $4$ &
$^{1}S$  &  $6$ & $(222)$ & $(00)$ & $3$  \\
$^{3}P$  &  $6$ & $(221)$ & $(11)$ & $4$ &
$^{3}H$  &  $6$ & $(221)$ & $(11)$ & $5$ &
$^{1}S$  &  $6$ & $(222)$ & $(40)$ & $4$  \\
$^{3}P$  &  $6$ & $(221)$ & $(30)$ & $5$ &
$^{3}H$  &  $6$ & $(221)$ & $(21)$ & $6$ &
$^{1}P$  &  $6$ & $(222)$ & $(30)$ & $0$  \\
$^{3}P$  &  $6$ & $(221)$ & $(31)$ & $6$ &
$^{3}H$  &  $6$ & $(221)$ & $(30)$ & $7$ &
$^{1}D$  &  $2$ & $(200)$ & $(20)$ & $1$  \\
$^{3}D$  &  $4$ & $(211)$ & $(20)$ & $1$ &
$^{3}H$  &  $6$ & $(221)$ & $(31)$ & $8$ &
$^{1}D$  &  $4$ & $(220)$ & $(20)$ & $2$  \\
$^{3}D$  &  $4$ & $(211)$ & $(21)$ & $2$ &
$^{3}H$  &  $6$ & $(221)$ & $(31)$ & $9$ &
$^{1}D$  &  $4$ & $(220)$ & $(21)$ & $3$  \\
$^{3}D$  &  $6$ & $(221)$ & $(20)$ & $3$ &
$^{3}I$  &  $4$ & $(211)$ & $(20)$ & $1$ &
$^{1}D$  &  $4$ & $(220)$ & $(22)$ & $4$  \\
$^{3}D$  &  $6$ & $(221)$ & $(21)$ & $4$ &
$^{3}I$  &  $4$ & $(211)$ & $(30)$ & $2$ &
$^{1}D$  &  $6$ & $(222)$ & $(20)$ & $5$  \\
$^{3}D$  &  $6$ & $(221)$ & $(31)$ & $5$ &
$^{3}I$  &  $6$ & $(221)$ & $(20)$ & $3$ &
$^{1}D$  &  $6$ & $(222)$ & $(40)$ & $6$  \\
$^{3}F$  &  $2$ & $(110)$ & $(10)$ & $1$ &
$^{3}I$  &  $6$ & $(221)$ & $(30)$ & $4$ &
$^{1}F$  &  $4$ & $(220)$ & $(21)$ & $1$  \\
$^{3}F$  &  $4$ & $(211)$ & $(10)$ & $2$ &
$^{3}I$  &  $6$ & $(221)$ & $(31)$ & $5$ &
$^{1}F$  &  $6$ & $(222)$ & $(10)$ & $2$  \\
$^{3}F$  &  $4$ & $(211)$ & $(21)$ & $3$ &
$^{3}I$  &  $6$ & $(221)$ & $(31)$ & $6$ &
$^{1}F$  &  $6$ & $(222)$ & $(30)$ & $3$  \\
$^{3}F$  &  $4$ & $(211)$ & $(30)$ & $4$ &
$^{3}K$  &  $4$ & $(211)$ & $(21)$ & $1$ &
$^{1}F$  &  $6$ & $(222)$ & $(40)$ & $4$  \\
$^{3}F$  &  $6$ & $(221)$ & $(10)$ & $5$ &
$^{3}K$  &  $4$ & $(211)$ & $(30)$ & $2$ &
$^{1}G$  &  $2$ & $(200)$ & $(20)$ & $1$  \\
$^{3}F$  &  $6$ & $(221)$ & $(21)$ & $6$ &
$^{3}K$  &  $6$ & $(221)$ & $(21)$ & $3$ &
$^{1}G$  &  $4$ & $(220)$ & $(20)$ & $2$  \\
$^{3}F$  &  $6$ & $(221)$ & $(30)$ & $7$ &
$^{3}K$  &  $6$ & $(221)$ & $(30)$ & $4$ &
$^{1}G$  &  $4$ & $(220)$ & $(21)$ & $3$  \\
$^{3}F$  &  $6$ & $(221)$ & $(31)$ & $8$ &
$^{3}K$  &  $6$ & $(221)$ & $(31)$ & $5$ &
$^{1}G$  &  $4$ & $(220)$ & $(22)$ & $4$  \\
$^{3}F$  &  $6$ & $(221)$ & $(31)$ & $9$ &
$^{3}K$  &  $6$ & $(221)$ & $(31)$ & $6$ &
$^{1}G$  &  $6$ & $(222)$ & $(20)$ & $5$  \\
$^{3}G$  &  $4$ & $(211)$ & $(20)$ & $1$ &
$^{3}L$  &  $4$ & $(211)$ & $(21)$ & $1$ &
$^{1}G$  &  $6$ & $(222)$ & $(30)$ & $6$  \\
$^{3}G$  &  $4$ & $(211)$ & $(21)$ & $2$ &
$^{3}L$  &  $6$ & $(221)$ & $(21)$ & $2$ &
$^{1}G$  &  $6$ & $(222)$ & $(40)$ & $7$  \\
$^{3}G$  &  $4$ & $(211)$ & $(30)$ & $3$ &
$^{3}L$  &  $6$ & $(221)$ & $(31)$ & $3$ &
$^{1}G$  &  $6$ & $(222)$ & $(40)$ & $8$  \\
$^{3}G$  &  $6$ & $(221)$ & $(20)$ & $4$ &
$^{3}M$  &  $4$ & $(211)$ & $(30)$ & $1$ &
$^{1}H$  &  $4$ & $(220)$ & $(21)$ & $1$  \\
$^{3}G$  &  $6$ & $(221)$ & $(21)$ & $5$ &
$^{3}M$  &  $6$ & $(221)$ & $(30)$ & $2$ &
$^{1}H$  &  $4$ & $(220)$ & $(22)$ & $2$  \\
$^{3}G$  &  $6$ & $(221)$ & $(30)$ & $6$ &
$^{3}M$  &  $6$ & $(221)$ & $(31)$ & $3$ &
$^{1}H$  &  $6$ & $(222)$ & $(30)$ & $3$  \\
$^{3}G$  &  $6$ & $(221)$ & $(31)$ & $7$ &
$^{3}N$  &  $6$ & $(221)$ & $(31)$ & $0$ &
$^{1}H$  &  $6$ & $(222)$ & $(40)$ & $4$  \\
$^{3}H$  &  $2$ & $(110)$ & $(11)$ & $1$ &
$^{3}O$  &  $6$ & $(221)$ & $(31)$ & $0$ &
$^{1}I$  &  $2$ & $(200)$ & $(20)$ & $1$  \\
\hline

\end{tabular}
\end{center}
\end{table}

\clearpage
% Lentele 1.4

\begin{table}
%\caption{Termu charakteristikos $f$- sluoksniui.}
\begin{center}
\begin{tabular}{ l c c c l | r c c c l | r c c c l }
\hline \hline
$LS$  & $\nu$ & $W$ & $U$ & $Nr$ &
$LS$  & $\nu$ & $W$ & $U$ & $Nr$ &
$LS$  & $\nu$ & $W$ & $U$ & $Nr$\\ \hline
$^{1}I$  &  $4$ & $(220)$ & $(20)$ & $2$ &
$^{4}P$  &  $5$ & $(211)$ & $(11)$ & $1$ &
$^{4}I$  &  $7$ & $(220)$ & $(20)$ & $4$  \\
$^{1}I$  &  $4$ & $(220)$ & $(22)$ & $3$ &
$^{4}P$  &  $5$ & $(211)$ & $(30)$ & $2$ &
$^{4}I$  &  $7$ & $(220)$ & $(22)$ & $5$  \\
$^{1}I$  &  $6$ & $(222)$ & $(20)$ & $4$ &
$^{4}D$  &  $3$ & $(111)$ & $(20)$ & $1$ &
$^{4}K$  &  $5$ & $(211)$ & $(21)$ & $1$  \\
$^{1}I$  &  $6$ & $(222)$ & $(30)$ & $5$ &
$^{4}D$  &  $5$ & $(211)$ & $(20)$ & $2$ &
$^{4}K$  &  $5$ & $(211)$ & $(30)$ & $2$  \\
$^{1}I$  &  $6$ & $(222)$ & $(40)$ & $6$ &
$^{4}D$  &  $5$ & $(211)$ & $(21)$ & $3$ &
$^{4}K$  &  $7$ & $(220)$ & $(21)$ & $3$  \\
$^{1}I$  &  $6$ & $(222)$ & $(40)$ & $7$ &
$^{4}D$  &  $7$ & $(220)$ & $(20)$ & $4$ &
$^{4}L$  &  $5$ & $(211)$ & $(21)$ & $1$  \\
$^{1}K$  &  $4$ & $(220)$ & $(21)$ & $1$ &
$^{4}D$  &  $7$ & $(220)$ & $(21)$ & $5$ &
$^{4}L$  &  $7$ & $(220)$ & $(21)$ & $2$  \\
$^{1}K$  &  $6$ & $(222)$ & $(30)$ & $2$ &
$^{4}D$  &  $7$ & $(220)$ & $(22)$ & $6$ &
$^{4}L$  &  $7$ & $(220)$ & $(22)$ & $3$  \\
$^{1}K$  &  $6$ & $(222)$ & $(40)$ & $3$ &
$^{4}F$  &  $3$ & $(111)$ & $(10)$ & $1$ &
$^{4}M$  &  $5$ & $(211)$ & $(30)$ & $0$  \\
$^{1}L$  &  $4$ & $(220)$ & $(21)$ & $1$ &
$^{4}F$  &  $5$ & $(211)$ & $(10)$ & $2$ &
$^{4}N$  &  $7$ & $(220)$ & $(22)$ & $0$  \\
$^{1}L$  &  $4$ & $(220)$ & $(22)$ & $2$ &
$^{4}F$  &  $5$ & $(221)$ & $(21)$ & $3$ &
$^{2}S$  &  $7$ & $(222)$ & $(00)$ & $1$  \\
$^{1}L$  &  $6$ & $(222)$ & $(40)$ & $3$ &
$^{4}F$  &  $5$ & $(211)$ & $(30)$ & $4$ &
$^{2}S$  &  $7$ & $(222)$ & $(40)$ & $2$  \\
$^{1}L$  &  $6$ & $(222)$ & $(40)$ & $4$ &
$^{4}F$  &  $7$ & $(220)$ & $(21)$ & $5$ &
$^{2}P$  &  $3$ & $(210)$ & $(11)$ & $1$  \\
$^{1}M$  &  $6$ & $(222)$ & $(30)$ & $1$ &
$^{4}G$  &  $3$ & $(111)$ & $(20)$ & $1$ &
$^{2}P$  &  $5$ & $(221)$ & $(11)$ & $2$  \\
$^{1}M$  &  $6$ & $(222)$ & $(40)$ & $2$ &
$^{4}G$  &  $5$ & $(211)$ & $(20)$ & $2$ &
$^{2}P$  &  $5$ & $(221)$ & $(30)$ & $3$  \\
$^{1}N$  &  $4$ & $(220)$ & $(22)$ & $1$ &
$^{4}G$  &  $5$ & $(211)$ & $(21)$ & $3$ &
$^{2}P$  &  $5$ & $(221)$ & $(31)$ & $4$  \\
$^{1}N$  &  $6$ & $(222)$ & $(40)$ & $2$ &
$^{4}G$  &  $5$ & $(211)$ & $(30)$ & $4$ &
$^{2}P$  &  $7$ & $(222)$ & $(30)$ & $5$  \\
$^{1}Q$  &  $6$ & $(222)$ & $(40)$ & $0$ &
$^{4}G$  &  $7$ & $(220)$ & $(20)$ & $5$ &
$^{2}D$  &  $3$ & $(210)$ & $(20)$ & $1$  \\
\multicolumn{4}{l}{\sl subshell $f^{7}$}&  &
$^{4}G$  &  $7$ & $(220)$ & $(21)$ & $6$ &
$^{2}D$  &  $3$ & $(210)$ & $(21)$ & $2$  \\
$^{8}S$  &  $7$ & $(000)$ & $(00)$ & $0$ &
$^{4}G$  &  $7$ & $(220)$ & $(22)$ & $7$ &
$^{2}D$  &  $5$ & $(221)$ & $(20)$ & $3$  \\
$^{6}P$  &  $5$ & $(110)$ & $(11)$ & $0$ &
$^{4}H$  &  $5$ & $(211)$ & $(11)$ & $1$ &
$^{2}D$  &  $5$ & $(221)$ & $(21)$ & $4$  \\
$^{6}D$  &  $7$ & $(200)$ & $(20)$ & $0$ &
$^{4}H$  &  $5$ & $(221)$ & $(21)$ & $2$ &
$^{2}D$  &  $5$ & $(221)$ & $(31)$ & $5$  \\
$^{6}F$  &  $5$ & $(110)$ & $(10)$ & $0$ &
$^{4}H$  &  $5$ & $(221)$ & $(30)$ & $3$ &
$^{2}D$  &  $7$ & $(222)$ & $(20)$ & $6$  \\
$^{6}G$  &  $7$ & $(200)$ & $(20)$ & $0$ &
$^{4}H$  &  $7$ & $(220)$ & $(21)$ & $4$ &
$^{2}D$  &  $7$ & $(222)$ & $(40)$ & $7$  \\
$^{6}H$  &  $5$ & $(110)$ & $(11)$ & $0$ &
$^{4}H$  &  $7$ & $(220)$ & $(22)$ & $5$ &
$^{2}F$  &  $1$ & $(100)$ & $(10)$ & $1$  \\
$^{6}I$  &  $7$ & $(200)$ & $(20)$ & $0$ &
$^{4}I$  &  $3$ & $(111)$ & $(20)$ & $1$ &
$^{2}F$  &  $3$ & $(210)$ & $(21)$ & $2$  \\
$^{4}S$  &  $3$ & $(111)$ & $(00)$ & $1$ &
$^{4}I$  &  $5$ & $(211)$ & $(20)$ & $2$ &
$^{2}F$  &  $5$ & $(221)$ & $(10)$ & $3$  \\
$^{4}S$  &  $7$ & $(220)$ & $(22)$ & $2$ &
$^{4}I$  &  $5$ & $(211)$ & $(30)$ & $3$ &
$^{2}F$  &  $5$ & $(221)$ & $(21)$ & $4$  \\
\hline

\end{tabular}
\end{center}
\end{table}

\clearpage
% Lentele 1.5

\begin{table}
%\caption{Termu charakteristikos $f$- sluoksniui.}
\begin{center}
\begin{tabular}{ l c c c l| r c c c l r c c c l }
\hline \hline
$LS$  & $\nu$ & $W$ & $U$ & $Nr$ &
$LS$  & $\nu$ & $W$ & $U$ & $Nr$\\ \hline
$^{2}F$  &  $5$ & $(221)$ & $(30)$ & $5$ &
$^{2}I$  &  $5$ & $(221)$ & $(30)$ & $3$  \\
$^{2}F$  &  $5$ & $(221)$ & $(31)$ & $6$ &
$^{2}I$  &  $5$ & $(221)$ & $(31)$ & $4$  \\
$^{2}F$  &  $5$ & $(221)$ & $(31)$ & $7$ &
$^{2}I$  &  $5$ & $(221)$ & $(31)$ & $5$  \\
$^{2}F$  &  $7$ & $(222)$ & $(10)$ & $8$ &
$^{2}I$  &  $7$ & $(222)$ & $(20)$ & $6$  \\
$^{2}F$  &  $7$ & $(222)$ & $(30)$ & $9$ &
$^{2}I$  &  $7$ & $(222)$ & $(30)$ & $7$  \\
$^{2}F$  &  $7$ & $(222)$ & $(40)$ & $A$ &
$^{2}I$  &  $7$ & $(222)$ & $(40)$ & $8$  \\
$^{2}G$  &  $3$ & $(210)$ & $(20)$ & $1$ &
$^{2}I$  &  $7$ & $(222)$ & $(40)$ & $9$  \\
$^{2}G$  &  $6$ & $(210)$ & $(21)$ & $2$ &
$^{2}K$  &  $3$ & $(210)$ & $(21)$ & $1$  \\
$^{2}G$  &  $5$ & $(221)$ & $(20)$ & $3$ &
$^{2}K$  &  $5$ & $(221)$ & $(21)$ & $2$  \\
$^{2}G$  &  $5$ & $(221)$ & $(21)$ & $4$ &
$^{2}K$  &  $5$ & $(221)$ & $(30)$ & $3$  \\
$^{2}G$  &  $5$ & $(221)$ & $(30)$ & $5$ &
$^{2}K$  &  $5$ & $(221)$ & $(31)$ & $4$  \\
$^{2}G$  &  $5$ & $(221)$ & $(31)$ & $6$ &
$^{2}K$  &  $5$ & $(221)$ & $(31)$ & $5$  \\
$^{2}G$  &  $7$ & $(222)$ & $(20)$ & $7$ &
$^{2}K$  &  $7$ & $(222)$ & $(30)$ & $6$  \\
$^{2}G$  &  $7$ & $(222)$ & $(30)$ & $8$ &
$^{2}K$  &  $7$ & $(222)$ & $(40)$ & $7$  \\
$^{2}G$  &  $7$ & $(222)$ & $(40)$ & $9$ &
$^{2}L$  &  $3$ & $(210)$ & $(21)$ & $1$  \\
$^{2}G$  &  $7$ & $(222)$ & $(40)$ & $A$ &
$^{2}L$  &  $5$ & $(221)$ & $(21)$ & $2$  \\
$^{2}H$  &  $3$ & $(210)$ & $(11)$ & $1$ &
$^{2}L$  &  $5$ & $(221)$ & $(31)$ & $3$  \\
$^{2}H$  &  $3$ & $(210)$ & $(21)$ & $2$ &
$^{2}L$  &  $7$ & $(222)$ & $(40)$ & $4$  \\
$^{2}H$  &  $5$ & $(221)$ & $(11)$ & $3$ &
$^{2}L$  &  $7$ & $(222)$ & $(40)$ & $5$  \\
$^{2}H$  &  $5$ & $(221)$ & $(21)$ & $4$ &
$^{2}M$  &  $5$ & $(221)$ & $(30)$ & $1$  \\
$^{2}H$  &  $5$ & $(211)$ & $(30)$ & $5$ &
$^{2}M$  &  $5$ & $(221)$ & $(31)$ & $2$  \\
$^{2}H$  &  $5$ & $(221)$ & $(31)$ & $6$ &
$^{2}M$  &  $7$ & $(222)$ & $(30)$ & $3$  \\
$^{2}H$  &  $5$ & $(221)$ & $(31)$ & $7$ &
$^{2}M$  &  $7$ & $(222)$ & $(40)$ & $4$  \\
$^{2}H$  &  $7$ & $(222)$ & $(30)$ & $8$ &
$^{2}N$  &  $5$ & $(221)$ & $(31)$ & $1$  \\
$^{2}H$  &  $7$ & $(222)$ & $(40)$ & $9$ &
$^{2}N$  &  $7$ & $(222)$ & $(40)$ & $2$  \\
$^{2}I$  &  $3$ & $(210)$ & $(20)$ & $1$ &
$^{2}O$  &  $5$ & $(221)$ & $(31)$ & $0$  \\
$^{2}I$  &  $5$ & $(221)$ & $(20)$ & $2$ &
$^{2}Q$  &  $7$ & $(222)$ & $(40)$ & $0$  \\
\hline

\end{tabular}
\end{center}
\end{table}

\clearpage

No initial estimates were provided to the program.
Non-physical situations may then arise in the SCF
process~\cite{HFF}.  The output shows one such situation during
the first iteration, but the program recovers and during
later iterasions such warnings
no longer appear.  The ratio of the potential and kinetic energy
is close to -2.0, as required by the virial theorem, showing
good convergence for the wave function.

\section*{Acknowledgements}

One of the authors (GG) would like to thank M.R. Godefroid for the
providing
the opportunity to perform the testing of the program at the Universit\'e
Libre de
Bruxelles (BELGIUM) and for valuable suggestions during
the course of development of this version of program.
This research is part of  co-operative research project funded
by the National Science Foundation under grant No. PHY-9501830,

%______________________________________

\clearpage

\newpage

\section*{TEST RUN OUTPUT}

\begin{scriptsize}
%\small
\begin{verbatim}

>>hf
                      =============================
                       H A R T R E E - F O C K . 96
                      =============================

               THE DIMENSIONS FOR THE CURRENT VERSION ARE:
                          NWF= 20        NO=220

  START OF CASE
  =============

  Enter ATOM,TERM,Z
  Examples: O,3P,8. or Oxygen,AV,8.
>Nd+,4F,60.
  List the CLOSED shells in the fields indicated (blank line if none)
  ... ... ... ... ... ... ... ... etc.
>  1s  2s  2p  3s  3p  3d  4s  4p  4d  5s  5p
  Enter electrons outside CLOSED shells (blank line if none)
  Example: 2s(1)2p(3)
>4f(4)6s(1)
  There are 13 orbitals as follows:
    1s  2s  2p  3s  3p  3d  4s  4p  4d  5s  5p  4f  6s
  Orbitals to be varied: ALL/NONE/=i (last i)/comma delimited list/H
>all
  Default electron parameters ? (Y/N/H)
>y
  Ambiguous l**n parent term: Enter term and Nr for f-subshells
>3F3
  Default values for remaining parameters? (Y/N/H)
>y

          WEAK ORTHOGONALIZATION DURING THE SCF CYCLE=   T
          SCF CONVERGENCE TOLERANCE (FUNCTIONS)      = 1.00D-08
          NUMBER OF POINTS IN THE MAXIMUM RANGE      = 220

          ITERATION NUMBER  1
          ----------------

          SCF CONVERGENCE CRITERIA (SCFTOL*SQRT(Z*NWF)) =   2.8D-07

          C( 1s 6s) =    -0.03084   V( 1s 6s) = -1565.84714   EPS = 0.000020
          C( 2s 6s) =    -0.06406   V( 2s 6s) =  -274.28004   EPS = 0.000234
          C( 3s 6s) =    -0.16604   V( 3s 6s) =   -72.19584   EPS = 0.002300
          C( 4s 6s) =     0.02646   V( 4s 6s) =   -10.07054   EPS =-0.002627
          C( 5s 6s) =     0.12630   V( 5s 6s) =    -0.97428   EPS =-0.129631
          E( 6s 1s) =     0.00000   E( 1s 6s) =     0.00000
          E( 6s 2s) =    -0.00002   E( 2s 6s) =    -0.00001
          E( 6s 3s) =    -0.00025   E( 3s 6s) =    -0.00012
          E( 6s 4s) =     0.00017   E( 4s 6s) =     0.00008
          E( 6s 5s) =     0.01744   E( 5s 6s) =     0.00872

                     EL         ED             AZ           NORM       DPM
                     1s   3128.6805851    921.2342946   1.0536970    4.19D-02
                     2s    548.0200388    304.9174534   1.0620778    6.60D-02
                     2p    522.7712426   5018.8746437   1.3675876    1.81D-01
                     3s    144.2704835    142.1548417   5.7746458    4.17D-01
                     3p    140.9608657   2469.2031309   0.9319004    4.25D-01
                     3d    119.7093213   8458.4713990   0.7705958    5.91D-01
                     4s     43.7551400     86.7032048   0.3218645    2.57D+00
                     4p     39.3455261   1332.3542693   0.6952517    2.38D+00
                     4d     26.7211399   4199.0102081   0.4802196    1.82D+00
                     4f      7.5878908   1305.5948122   0.5721312    9.82D-01
                     5s      6.3617680     27.3266729   1.0694868    8.96D-01
                     5p      4.0723277    429.0689970   0.9602206    7.93D-01
          ED =   0.029648; ADJUSTED TO ALLOWED MINIMUM ENERGY
                     6s      0.4200626      8.7147854   0.2148108    4.65D-01
                     4p     14.0173622   1022.4876672   2.1997080    3.93D-01
                     4d      8.3655861   3338.2783677   1.6493227    2.70D-01
                     4s     20.1101311     65.3845359   8.0426513    5.70D-01
                     4f      1.9641284   1955.2655220   1.1526275    1.99D-01
                     5p      2.6937391    310.8797108   0.8665884    3.37D-01

      < 1s| 2s>= 2.4D-04
      < 1s| 3s>= 1.2D-03
      < 2s| 3s>= 1.0D-02
      < 2p| 3p>= 3.4D-03
      < 1s| 4s>= 5.3D-04
      < 2s| 4s>= 8.0D-03
      < 3s| 4s>= 6.0D-02
      < 2p| 4p>= 6.3D-03
      < 3p| 4p>= 7.4D-02
      < 3d| 4d>= 6.5D-02
      < 1s| 5s>= 1.4D-04
      < 2s| 5s>= 3.2D-03
      < 3s| 5s>= 2.4D-02
      < 4s| 5s>= 5.6D-02
      < 2p| 5p>= 4.2D-04
      < 3p| 5p>= 1.1D-03
      < 4p| 5p>=-1.4D-01
      < 1s| 6s>=-3.7D-05
      < 2s| 6s>= 8.0D-04
      < 3s| 6s>= 1.0D-02
      < 4s| 6s>= 5.6D-02
      < 5s| 6s>= 4.0D-01

    ... Iterations Omitted for Brevity ...

          ITERATION NUMBER  8
          ----------------

          SCF CONVERGENCE CRITERIA (SCFTOL*SQRT(Z*NWF)) =   3.6D-05

          C( 1s 6s) =     0.00000   V( 1s 6s) = -1527.02399   EPS = 0.000000
          C( 2s 6s) =     0.00000   V( 2s 6s) =  -243.19942   EPS = 0.000000
          C( 3s 6s) =     0.00000   V( 3s 6s) =   -54.06456   EPS = 0.000000
          C( 4s 6s) =     0.00000   V( 4s 6s) =   -11.16158   EPS = 0.000000
          C( 5s 6s) =     0.00000   V( 5s 6s) =    -1.49103   EPS = 0.000000
          E( 6s 1s) =     0.00001   E( 1s 6s) =     0.00001
          E( 6s 2s) =     0.00004   E( 2s 6s) =     0.00002
          E( 6s 3s) =    -0.00348   E( 3s 6s) =    -0.00174
          E( 6s 4s) =    -0.00227   E( 4s 6s) =    -0.00113
          E( 6s 5s) =     0.02248   E( 5s 6s) =     0.01124

                     EL         ED             AZ           NORM       DPM
                     1s   3054.8943660    921.1231536   1.0000000    5.06D-09
                     2s    487.2430204    301.6764224   1.0000000    1.19D-07
                     2p    460.9251144   4957.8113013   1.0000000    1.18D-07
                     3s    109.0072008    137.6751878   0.9999999    4.08D-07
                     3p     97.3900098   2374.4492227   0.9999999    4.38D-07
                     3d     76.0249753   7959.4441188   0.9999999    3.39D-07
                     4s     23.2551364     65.1851169   1.0000000    6.20D-07
                     4p     18.6793842   1101.2565739   1.0000000    6.83D-07
                     4d     10.5291882   3652.0923901   1.0000002    4.94D-07
                     4f      1.4033321   1922.9003308   1.0000011    1.71D-06
                     5s      3.8125602     25.1175225   1.0000001    6.86D-07
                     5p      2.4248720    387.7133668   1.0000000    7.49D-07
                     6s      0.7066424      7.8763382   0.9999988    9.91D-07

      < 1s| 2s>= 7.2D-10
      < 1s| 3s>= 4.8D-10
      < 2s| 3s>=-1.6D-09
      < 2p| 3p>=-6.9D-10
      < 1s| 4s>= 1.8D-10
      < 2s| 4s>= 1.8D-09
      < 3s| 4s>=-1.2D-08
      < 2p| 4p>= 1.6D-09
      < 3p| 4p>=-7.7D-09
      < 3d| 4d>=-1.5D-08
      < 3s| 5s>=-8.5D-09
      < 4s| 5s>=-2.1D-07
      < 2p| 5p>=-1.2D-10
      < 3p| 5p>=-9.8D-09
      < 4p| 5p>=-2.2D-07
      < 2s| 6s>= 1.3D-10
      < 3s| 6s>= 2.1D-10
      < 4s| 6s>=-2.9D-08
      < 5s| 6s>=-1.4D-07

     TOTAL ENERGY (a.u.)
     ----- ------
           Non-Relativistic    -9283.50837583    Kinetic     9283.50834626
           Relativistic Shift   -305.34173297    Potential -18567.01672209
           Relativistic        -9588.85010880    Ratio        -2.000000003

  Additional parameters ? (Y/N/H)
>n
  Do you wish to continue along the sequence ?
>n

  END OF CASE
  ===========
\end{verbatim}

\end{scriptsize}


\begin{thebibliography}{10}

\bibitem{GRASP} F. Parpia, C. Froese Fischer, I. Grant, Computer
Phys. Commun. (in press).

\bibitem{Tatewaki} H. Tatewaki, M. Sekiya, F. Sasaki,
O. Matsuoka, and T. Koga, Phys. Rev. A {\bf 51}, 197 (1995).

\bibitem{FF}C. Froese Fischer, Comput. Phys. Commun. 43 (1987) 355.

\bibitem{S} John C. Slater, Quantum Theory of Atomic Structure, Vol II.
(McGraw-Hill Book Company, New York, 1960), p. 439.

\bibitem{atsp} C. Froese Fischer, Comput. Phys. Commun. {\bf
64} (1991) 369.

\bibitem{HFF} C. Froese Fischer, The Hartree-Fock Method for Atoms
              (Wiley, New york, 1977).

\bibitem{NK} C.W. Nielson, G. Koster, Spectroscopic Coefficients for the
$p^{n}$, $d^{n}$, and $f^{n}$ Configurations (The M.I.T. Press, Cambridge,
1963).
\end{thebibliography}
\end{document}